\documentstyle[12pt,epsf]{article}

\advance \textheight by \topskip
\begin{document}

\title{Primordial Black Hole Binaries as a Source of Gamma-Ray
Bursts and of a High-Frequent Gravitational Radiation}

\author{J. N. Abdurashitov, V. E. Yants\\
\it Institute for Nuclear Research, Russian Academy of Sciences,\\
\it 117312 Moscow, Russia\\
C. V. Parfenov\\
\it Moscow State University, 119899 Moscow, Russia}

\maketitle

\begin{abstract}
Ultracompact primordial black hole binaries (PBHB's) with
masses $m > 10^{16} g$ are considered here. If PBHB's
contribute significant part of the dark matter of the Galaxy
one can expect an existence of high-frequent non-thermal
diffuse gravitational radiation with flux of $~1\ erg\ cm^{-2}\
s^{-1}$. The possibility of coalescence of the PBHB's in
Galaxy's halo to be a source at least of a part of gamma-ray
bursts (GRB) observed is discussed.
The energy flux of gravitational radiation from those GRB
should exceed the energy flux of $\gamma$-radiation by
7-8 orders of magnitude. The possibility of observation
of PBHB through detection of the gravitational radiation
burst coincident with GRB is emphasized. The PBHB also can be
observed detecting a stationary gravitatonal radiation
in the frequency range $> 10^4\ Hz$ and observing a high-frequent
pulsation of a source's brightness in microlensing effects
in the Galaxy's halo.
\end{abstract}
\newpage
An existence of primordial black holes (PBH) was predicted in
\cite{pha}. As it is known PBH may burn in many models of
evolution of early Universe \cite{haw,bkt,jed,bul}, and the
initial mass distribution is strongly model-dependent
\cite{bul, mcg}. More
often it appears to be close to $n(m)\sim{m^{-5/2}}$, but some
models suggest extended number of PBH with $m>10^{17}g$ - it
allows one to consider PBH as a main source of the dark matter
of the Universe \cite{mcg}. In the last years a possibility of existence
of the PBH with masses up to $0.5 M _{\odot}$ is intensively
discussed - these PBH are the best candidates as a lensing
bodies in the halo of Galaxy. PBH with $m<10^{15}g$ must
evaporate before our epoch. Nevertheless, assuming distributions
of PBH and barionic (visible) matter in the Universe to correlate
one can predict the modern density of PBH in the Galaxy to be more
or less high. If $\eta$ is a part of PBH with an average mass
$\overline{M}$ in the density of dark matter then average distance
between them is ($H\equiv H_0/100 s^{-1}Mpc^{-1}$ - modern
Hubble's constant):
\begin{equation}
\label{avg-dis}
\overline{r} \sim 4{\cdot}10^{18}\left(\frac{\overline{M}}{\eta
M_{\odot}}\right)^{1/3}(\Omega H^2)^{-4/3}\ cm
\end{equation}
Moreover, a capture of PBH by Solar system may result in
increase of PBH's concentration in the vicinity of the Sun.
In this case PBH may appear more closer to the Earth.

At final stage of evolution ($m<10^{14}$) when hadron burning
is possible the evaporation of PBH looks like an explosion observed
at far distance as burst of $\gamma$-rays in the energy range
$E_{\gamma}\sim0.1-100$ GeV during $\tau\sim10-50$ ms \cite{cli}
relating to some part of observed GRB's. Note, that the energy,
the duration and the spectrum of the burst created by final
explosion of PBH are unique and fixed, so most of GRB's which
are very manifold cannot be reasoned by this mechanism. Besides,
a substantial difficulty of evaporation model is that if all
observed GRB's are due to PBH explosions then $\gamma$-luminocity
of the Galaxy appears to be extremely high - in contradiction
with observed one.

Usually PBH is considered as a single object. However, many
astrophysical objects are observed to be a binary system.
Calculations show \cite{nak} that at the epoch of matter-radiation
equilibrium significant amount ($\kappa\sim0.05-0.03$) of PBH
can be captured into ultracompact binary systems due
to their very close disposition. Accordingly to \cite{nak} after
$T\sim10^{10}$ years the part of binary systems with lifetime
(before coalescence) $\tau\ll T$ is:

\begin{equation}
\label{par-lft}
p(T,\tau) \simeq 0.04\frac{\tau}{T}{\left(\frac{\overline{M}}
{\eta M_{\odot}}\right)}^{\frac{5}{148}}
\end{equation}

For example, when $\eta=1$, $\kappa=0.1$, $M_{\odot}=10^{18} g$,
$\tau=10^6 y$ then average distance between the Earth and nearest
binary PBH is $\sim10^{14} cm$. So it can be interesting to
search observable effects caused by existence of such objects
at same distances ($10^{14}-10^{16} cm)$.

If the distance between partners in a binary system is significantly
greater than their gravitational radii (that is their relative motion
is quasinewtonian) then for far observer it looks mainly like a source
of a high-frequent gravitational radiation with the intensity \cite{nak}

\begin{equation}
\label{int-gra}
I_g=\frac{32c^5}{5G}\left(\frac{M}{a}\right)^5f(\epsilon)
\end{equation}

Here $M\equiv\frac{G}{c^2}m_1m_2/(m_1+m_2)$ is gravitational radius, and
$f(\epsilon)\equiv(1-{\epsilon}^2)^{-7/2}(1+\frac{73}{24}{\epsilon}^2+
\frac{37}{96}{\epsilon}^4), a$ and $\epsilon$ are parameters of the orbit.
The intensity is determined by $\frac{M}{a}$ mainly, therefore at certain
stage of evolution of the binary system it can appear close to the intensity
of radiation of usual astrophysical binary objects. The main harmony in the
spectrum is

\begin{equation}
\label{fre-gra}
\omega=2\Omega_{rot}=\frac{2c}{a}\sqrt{\frac{m_1+m_2}{a}}
\end{equation}

and one should note, that the relative intensity of higher frequencies
increases with the excentricity $\epsilon$ increasing. The energy loose
due to gravitational waves emission leads PBH to fall each to other
during

\begin{equation}
\label{tim-fal}
\tau=t_{fall}-t_0\simeq\frac{5}{256c}\frac{a_0^4}{M(m_1+m_2)^2}
\end{equation}

Comparing (\ref{tim-fal}) and (\ref{int-gra}) one concludes that PBH binary
emits significantly only at very late stage of evolution. For example,
when $m_1=m_2=10^{17} g$ and $\frac{a_0}{M}<10^9$ then intensity
$I>10^{10} Wt$ and $\tau<5\cdot10^3$ years.

At distance $r$ from binary system amplitude of gravitational
wave is

\begin{equation}
\label{amp-gra}
h\sim\frac{M^2}{ar}
\end{equation}

\begin{figure}
\begin{centering}
\epsfsize=0.6\textwidth
\epsffile{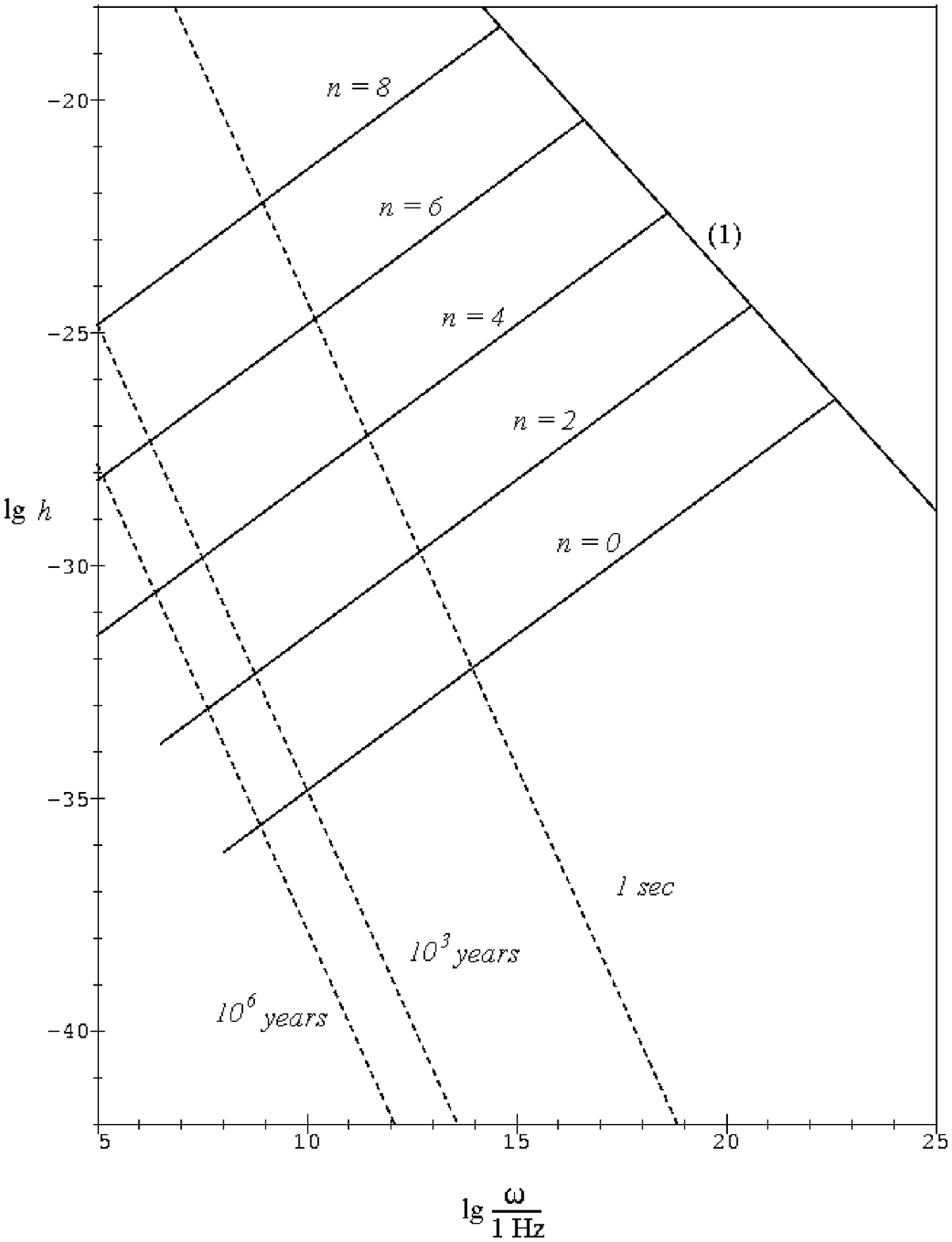}
\caption{Lines of PBH binary evolution}
\label{hoe}
\end{centering}
\end{figure}

Therefore, at $h-\omega$ plot the point of radiation which can
be observed near the Earth shifts to a high $h$ and $\omega$ region
during evolution of the PBH binary, thus moving to a coalescense
line. In the Fig.\ \ref{hoe} the lines of evolution of the radiation from
symmetric ($m_1=m_2=10^{16+n} g$) at distance $10^{14} cm$
are plotted. Dotted lines express constant time left until
coalescense.

At the moment of coalescense almost all energy is emitted
as a short powerful burst of gravitational waves.
This process cannot be described
in the framework of quasinewtonian approach. To estimate
amplitude and duration of the burst one usually
extrapolates the problem of a probe body fall into
a black hole. As a PBH binary has large orbital momentum
and high rotational velocity, so when it merges one can
expect the efficiency of emission to be high enough
$\xi=E_g/(m_1+m_2)c^2\sim0.06-0.2$, closing
to thermodynamical limit $\xi_{max}=1-1/\sqrt{2}$.
Thus, at the end of existence ultracompact PBH binary
emits short
(calculations give $\delta{t}\sim(10^1-10^3)\frac{M}{c}$)
powerful ($E_g\sim0.1(m_1+m_2)c^2$) burst of gravitational
radiation. As a result the total gravitational radiation
luminocity due to PBH binaries coalescense in the Galaxy
where $M_{DM}\simeq50M_{Galaxy}$ is

\begin{equation}
\label{int_gal}
I_G\simeq{E_g}(\kappa\eta{M_{DM}}/\overline{M}
\cdot(P(T,\tau)/\tau)
\simeq0.001\kappa\eta{M_{DM}}c^2/T\sim\kappa\eta10^{39}\ Wt
\end{equation}

Therefore, if PBH binaries exist then the diffuse flux of
non-thermal high-frequent ($\omega>10^{10}Hz$) gravitational
radiation should be there in the Galaxy, which has
the density

\begin{equation}
\label{dif-gra}
S_G\simeq{I_{G}}/4{\pi}{R_G}^2\sim\kappa\eta\cdot0.5\ erg\ cm^{-1}\ s^{-1}
\end{equation}

Other possible effect of PBH binaries is an intensive particles creation
and production of a fireball due to the energy being released during
the merging. Since the amount of the energy is relatively large,
the intensity of neutrino and $\gamma$-radiation can appear to be
significant too even with a very small conversion efficiency.
The part of energy $\epsilon_{\gamma}$ being converted
to $\gamma$-radiation cannot exceed the value of $10^{-8}/\kappa\eta$,
which provides almost all $\gamma$-luminocity of the Galaxy
($\sim10^{31}$ W) assuming the dark matter to consist of PBH binaries.
Besides, if parthers of PBH binaries have a mass $m\gg10^{16}\ g$
the $\gamma$-evaporation is very small, and no problem of extremely
high background of the diffuse $\gamma$-radiation is appeared.

At far distance the coalescense can be observed as neutrino and
$\gamma$-bursts emited by relativistic fireball. The specific feature
of this should be a homogeneous distribution in the Galaxy and hence
a characteristic dependence of the number of sources on the flux observed
above threshold ($\lg{N} \propto -3/2 \lg{S}$). But the total range of
observed $\gamma$-bursts has not such distribution, i.e. the dependence
$\lg{N}-\lg{S}$ deviates from simple 3/2 law. Moreover, several GRB's
are observed to have extremely high redshift during optical counterpart,
i.e. some of GRB's observed are obviously cosmological \cite{pir}.
Therefore one cannot consider PBH binaries as unique source of GRB.
Note, that the GRB's are divided on several classes. There are both
relatively short bursts with $\Delta{t}_\gamma < 1 s$ and long one
with $\Delta{t}_\gamma > 10 s$, and all GRB are widely distributed
by hardness of spectrum.(ref to catalog to be here?)
Besides, only long GRB with hard spectrum have significant deviation
from -3/2 law. This can lead one to conclude that the origin of
different GRB can differ too, in particular, significant
part of GRB can burn in the Galaxy's halo. Thus, it's possible
to suggest the fireball burning due to PBH merging in the halo
to be a source of some part of GRB's. The merging rate
in the halo can be estimated as

\begin{equation}
\label{mrg-rat}
\nu\simeq(\kappa\eta M_{DM}/\overline{M})(P(T,\tau)/\tau)
\simeq10^{-6}\ \kappa\eta(M_{\odot}/\overline{M})\ s^{-1}
\end{equation}

The rate of the GRB which can be observed near the Earth
is determined by an observational threshold $S_{min}$ and
an average mass of halo's PBH:
$\nu_{GRB}\simeq\nu(r/r_G)^3$, where
$r\simeq(\epsilon_{\gamma}\cdot0.1\overline{M}c^2/4\pi
S_{min})^{1/2}$, and therefore

\begin{equation}
\label{grb-rat}
\nu_{GRB}\simeq5\cdot10^2\ s^{-1}
\kappa\eta(\epsilon_{\gamma}/10^{-8})^{3/2}
(\overline{M}/M_{\odot})^{1/2}
(10^{-7}\ erg\ cm^{-1}\ s^{-1}/S_{min})^{3/2}
\end{equation}

Taking into account that the rate of euclidian GRB's
above $S_{min}=10^{-7}\ erg\ cm^{-1}\ s^{-1}$
is $\sim10^{-5}\ s^{-1}$ one can obtain a lower limit
on average mass of PBH in the halo. For example,
if $\eta=1,\ \kappa=1$ and $\epsilon_{\gamma}=10^{-8}$
then $\overline{M}\ge10^{18}\ $g.

In frame of the model, as the $\gamma$-luminocity
of the Galaxy is limited ($\epsilon_{\gamma}<10^{-8}$) so
the energy of gravitational radiation outburst from PBH binary
coalescence may exceed the $\gamma$-burst energy by 8 orders
of magnitude, that is, the gavitational energy fluence observed
near the Earth during powerful GRB can be greater than
$10^4\ erg\ cm^{-2}$.

One should note, that the discovery of very characteristic
high-frequent gravitational radiation would prove
an existence of ultracompact PBH binaries.
But all existing and building gravitational detectors are oriented
toward the search of low-frequent one \cite{rdg}, and so
are not able to detect PBH binary through it's stationary
radiation. Therefore, it looks very actual to develop
the technique of detection of short intensive
bursts and stationary high-frequent gravitational
radiation like electromagnetic or plasma detectors \cite{rud, gri}.
For example, as it is shown at \cite{gri}, if one uses
electromagnetic field in the resonator with $Q=10^{12}$
as an antenna one can detect gravitational wave down to
$h\sim10^{-27}$ at frequency $10^9$\ Hz. But one should
note that the probability of detection of stationary
radiation with so high frequency is extremely low
due to short life time of compact binaries at this
frequency (from $10^6$\ years for
$m_1=m_2=10^{16}$\ g to 1\ year for $m_1=m_2=10^{20}$\ g).

In principle, when one observes a microlensing effect
with the binary consisting of the same mass PBH's
appeared to be a gravitational lens, one should observe
characteristic periodic pulsations of brightness of
the source. The magnitude of the effect is estimated to be
$\sim{a^2}/{b^2}$, where $b$ is the distance of the center
of binary masses from the direct line of sight, and
the frequency of pulsations is equal to $2\Omega_{rot}$.
Again, one should note, that the earth-based instruments
are not able to detect such pulsations due to influence
of atmosphere.

We would like to thank V. A. Rubakov for useful discussions.

\newpage

\end{document}